\documentclass[prd,superscriptaddress,amsfonts,amssymb,amsmath,showpacs,twocolumn]{revtex4}

\usepackage{amsmath}
\usepackage{epsfig}
\usepackage{graphics}
\usepackage{color}
\usepackage{graphicx}
\usepackage[font={footnotesize,it}]{caption}
\usepackage[colorlinks=true,
            linkcolor=red,
            urlcolor=blue,
            citecolor=green]{hyperref}

\begin{document}

\title{Deflection Angle of Light by Wormholes using the  Gauss-Bonnet Theorem}
\author{Kimet Jusufi}
\email{kimet.jusufi@unite.edu.mk}
\affiliation{Physics Department, State University of Tetovo, Ilinden Street nn, 1200,
Tetovo, Macedonia}
\affiliation{Institute of Physics, Faculty of Natural Sciences and Mathematics, Ss. Cyril and Methodius University, Arhimedova 3, 1000 Skopje, Macedonia}

\date{\today }

\begin{abstract}
In this letter, we have investigated the deflection angle of light by wormholes using a new geometrical method known as Gibbons--Werner method (GW). In particular we have calculated the deflection angle of light in the weak limit approximation in two wormhole spacetime geometries: Ellis wormhole and Janis--Newman--Winnicour (JNW) wormhole. We have employed the famous Gauss--Bonnet theorem (GBT) to the Ellis wormhole optical geometry and JNW wormhole optical geometry, respectively. By using GBT, we computed the deflection angles in leading orders by these wormholes and our results were compared with the ones
in the literature.
\end{abstract}

\pacs{04.40.-b, 95.30.Sf, 98.62.Sb}
\keywords{Light deflection; Janis--Newman--Winnicour Wormhole; Ellis Wormhole; Gauss-Bonnet Theorem}
\maketitle

\section{Introduction}

Einstein's general theory of relativity (GR) is the theoretical basis of modern physics which describes gravitation as a geometric theory of space-time.  Many theoretical predictions of Einstein's theory have been experimentally confirmed over the years. Gravitational lensing is a famous prediction by GR which has been confirmed in the early days during a total solar eclipse by Eddington and widely studied in modern observational cosmology \cite{sch,glensing}. On the other hand, wormholes arise as solutions to the Einstein's field equations of general relativity with no experimental evidence whatsoever until today \cite{tw1,tw2,tw3}. Today, the interest in those exotic objects continues to be of pure speculative nature.

In a recent study, Gibbons and Werner (GW), discovered a new method to calculate the deflection angle of light by emplying the GBT to the optical geometry \cite%
{gibbons1}. Furthermore they have shown this method to give an exact result in the weak limit approximation for Schwarzschild black hole spacetime \cite{gibbons1}. Then, this method was extended to calculate the deflection angle in a Kerr black hole \cite{werner}, rotating global monopole and cosmic string \cite{kimet1,kimet2,kimet3}, light deflection under the Lorentz symmetry breaking and  quantum effects \cite{kimet4,kimet5}, deflection of light in Rindler modified Schwarzschild black hole \cite{sakalli}, strong deflection limit for finite-distance corrections \cite{asahi1,asahi2,asahi3}, and references therein. The GW method is shown to be exact in leading order terms which is quite an amazing result since allows the computation of the deflection angle by choosing
a domain \textit{outside} of the light ray which is a bit different from the usual belief which describes light deflection as an effect related to the spacetime curvature due to the mass inside of the impact parameter.

Recently, gravitational lensing was also investigated in the context of wormholes. The deflection angle in Ellis geometry was first pointed out by Chetouani and Clement \cite{wh0}. More recently, Nakajima and Asada, studied light deflection by Ellis wormholes \cite{wh4}, morover gravitational microlensing by Ellis wormhole was investigated by Abe \cite{wh5}. Furthermore, Dey and Sen, studied the strong limit and the relativistic images in Janis--Newman--Winnicour (JNW) wormhole spacetime \cite{wh2}.  Nandi, Zhang and Zakharov studued the deflection of light by  wormhole solutions considered in the Einstein minimally coupled theory and in the brane world model \cite{wh3}. Bhattachary and Potapov investigated the deflection of light by Ellis wormhole using three different methods: direct integration method, perturbation method, and invariant angle method \cite{wh1}, whereas in Ref. \cite{wh6} wave effect in gravitational lensing by the Ellis Wormhole was investigated. Tsukamoto studied the strong and weak deflection limit by Ellis wormhole \cite{wh8,wh9,wh10}. In particular the deflection angle in the weak limit by Ellis wormhole was found to be \cite{wh1,wh4}
\begin{equation}
\hat{\alpha}\simeq \frac{\pi a^2}{4 b^2}+\mathcal{O}(\frac{a^4}{b^4}).  \label{1}
\end{equation}
where $b$ is the impact parameter. 

In this paper, we shall try to answer  the following question: can we calculate the deflection angle by wormholes using the GBT? More precisely, we shall elaborate two well known wormhole solutions: firstly, the Ellis wormhole, and secondly, the JNW wormhole. Let as note that, at large distance, the deflection angle by JNW wormhole using stadard methods was found to be \cite{wh2}
\begin{equation}
\hat{\alpha}\simeq \frac{4m \gamma}{r_{0}}+\mathcal{O}(\frac{m^2}{r_{0}^2}).  \label{2}
\end{equation}
where $\gamma>1$ corresponds to the (JNW) wormhole and $r_0$ is the distance of closest approach of light to the compact object which can be approximated with the impact parameter in the weak limit.

The paper is organized as follows. In Sec. II, we shall introduce the Ellis optical wormhole spacetime and calculate the Gaussian optical curvature. In the second part, we shall apply the GBT to this geometry to compute the deflection angle.  In Sec. III, we elborate on the JNW wormhole spacetime.  We draw our conclusions in Sec. IV. Throughout this paper we  shall use natural units, i.e. $G=c=\hbar=1$.

\section{Ellis wormhole}
\subsection{Gaussian Optical Curvature}
Let us start by writing the line element for the Ellis wormhole which is given by  \cite{wh4}
\begin{equation}
\mathrm{d}s^2=-\mathrm{d}t^2+\mathrm{d}r^2+(r^2+a^2)(\mathrm{d}\theta^2+\sin^2\theta \mathrm{d}\varphi^2),  \label{3}
\end{equation}
as was noted in  \cite{wh4}, in order to cover the entire spacetime geometry, the radial coordinate $r$ must run from $-\infty$ to $\infty$. In the special case, by setting $r=0$, the throat of the wormhole is recoverd. We shall restrict ourselves by considering only one part of the wormhole geometry by setting $r>0$. 

Let us now find the optical metric by considering first the null case by writing $\mathrm{d}s^{2}=0$, and after considering the deflection on the equatorial plane $\theta =\pi /2$, one can easily find the Ellis wormhole optical metric given by
\begin{equation}
\mathrm{d}t^{2}=\mathrm{d}r^{2}+(r^2+a^2)\mathrm{d}\varphi^2. \label{4}
\end{equation}%

After we introduce a new variables $r^{\star }$, and hence a new function $%
f(r^{\star })$, as follows
\begin{equation}
\mathrm{d}r^{\star }=\mathrm{d}r,\,\,\,f(r^{\star })=\sqrt{r^2+a^2 }, \label{5}
\end{equation}%
the Ellis wormhole optical metric reads
\begin{equation}
\mathrm{d}t^{2}=\tilde{g}_{ab}\,\mathrm{d}x^{a}\mathrm{d}x^{b}=\mathrm{d}{%
r^{\star }}^{2}+f^{2}(r^{\star })\mathrm{d}\varphi ^{2}. \,\,\,\,(a,b=r,\varphi) \label{6}
\end{equation}%

The nonzero components are found to be 
\begin{eqnarray}
\tilde{\Gamma} _{\varphi \varphi }^{r}&=&-r,\\
\tilde{\Gamma} _{r\varphi }^{\varphi}&=&\frac{r}{r^2+a^2}.
\end{eqnarray} 

The first important quantity to be found is the Gaussian optical curvature $K$. Following the arguments in Ref. \cite{gibbons1}, $K$ can be calculated as 
\begin{eqnarray}\label{9}
K &=&-\frac{1}{f(r^{\star })}\frac{\mathrm{d}^{2}f(r^{\star })}{\mathrm{d}{%
r^{\star }}^{2}},  \\
&=&-\frac{1}{f(r^{\star })}\left[ \frac{\mathrm{d}r}{\mathrm{d}r^{\star }}%
\frac{\mathrm{d}}{\mathrm{d}r}\left( \frac{\mathrm{d}r}{\mathrm{d}r^{\star }}%
\right) \frac{\mathrm{d}f}{\mathrm{d}r}+\left( \frac{\mathrm{d}r}{\mathrm{d}%
r^{\star }}\right) ^{2}\frac{\mathrm{d}^{2}f}{\mathrm{d}r^{2}}\right] .\notag
\end{eqnarray}%

Now taking into account Eqs.  \eqref{5} and \eqref{6},  after some algebra from Eq. \eqref{9} we obtain the following result
\begin{equation}\label{10}
K=-\frac{a^2}{\left(r^2+a^2\right)^2}
\end{equation}

Later on, we shall use this importan result together with the GBT to find
the deflection angle in the following section.

\subsection{Deflection angle by Ellis wormhole}

Having computed the Gaussian optical curvature we can proceed further by writing the GBT to the wormhole optical geometry More specifically, for the non-singular region $\mathcal{D}_{R}$ and boundary $\partial
\mathcal{D}_{R}=\gamma _{\tilde{g}}\cup C_{R}$ the GBT can be stated as follows \cite{gibbons1}
\begin{equation}
\iint\limits_{\mathcal{D}_{R}}K\,\mathrm{d}S+\oint\limits_{\partial \mathcal{%
D}_{R}}\kappa \,\mathrm{d}t+\sum_{i}\theta _{i}=2\pi \chi (\mathcal{D}_{R}).
\label{11}
\end{equation}

In the left hand side of the last equation $\kappa $ stands for the geodesic curvature, $K$ as we know stands for the Gaussian optical curvature, while $\theta_{i}$ is the exterior angle at the $i^{th}$ vertex. On the other hand  $%
\chi (\mathcal{D}_{R})$ is known as the Euler characteristic number. 

Furthermore we shall find the geodesic curvature which can be calculated as $\kappa =\tilde{g}\,(\nabla _{\dot{%
\gamma}}\dot{\gamma},\ddot{\gamma})$, with the unit speed relation given by $\tilde{g}(\dot{\gamma},\dot{%
\gamma})=1$, where $\ddot{\gamma}$ is the unit acceleration vector. In our setup, we are interested in the weak limit approximation, therefore if we let  $R\rightarrow \infty $, our two jump angles ($\theta _{\mathit{O}}$ , $%
\theta _{\mathit{S}}$) become $\pi /2,$. In other words the total angle is given as a sum of jump angle to the source $\mathit{S}$, and observer $\mathit{O}$, i.e. $\theta _{\mathit{O}%
}+\theta _{\mathit{S}}\rightarrow \pi $ \cite{gibbons1}.  The integration domain is considered to be outside of the light ray with the Euler characteristc number $\chi (\mathcal{D}_{R})=1$. The GBT yields
\begin{equation}
\iint\limits_{\mathcal{D}_{R}}K\,\mathrm{d}S+\oint\limits_{C_{R}}\kappa \,%
\mathrm{d}t\overset{{R\rightarrow \infty }}{=}\iint\limits_{\mathcal{D}%
_{\infty }}K\,\mathrm{d}S+\int\limits_{0}^{\pi +\hat{\alpha}}\mathrm{d}\varphi
=\pi
\end{equation}

Let us know compute the geodesic curvature $\kappa$. To do so, we first point out that $\kappa (\gamma _{\tilde{g}})=0$ since  $\gamma _{\tilde{g}}$ is a
geodesic.  We are left with the following $\kappa (C_{R})=|\nabla _{%
\dot{C}_{R}}\dot{C}_{R}|$,  where we have choosen $C_{R}:=r(\varphi
)=R=\text{const}$.  The radial part is evaluated as
\begin{equation}
\left( \nabla _{\dot{C}_{R}}\dot{C}_{R}\right) ^{r}=\dot{C}_{R}^{\varphi
}\,\left( \partial _{\varphi }\dot{C}_{R}^{r}\right) +\tilde{\Gamma} _{\varphi
\varphi }^{r}\left( \dot{C}_{R}^{\varphi }\right) ^{2}.  \label{12}
\end{equation}

The first term gives zero contribution and the second part can be calculated if we make use of the  $\tilde{\Gamma} _{\varphi \varphi }^{r}=-R$ and $\dot{C}_{R}^{\varphi }=1/\sqrt{R^2+a^2}$, following from the unit speed condition. Hence, the geodesic curvature  yields
\begin{eqnarray}\notag
\lim_{R\rightarrow \infty }\kappa (C_{R}) &=&\lim_{R\rightarrow \infty
}\left\vert \nabla _{\dot{C}_{R}}\dot{C}_{R}\right\vert ,  \notag \\
&=&\lim_{R\rightarrow \infty }\left( \frac{R}{R^2+a^2}\right),  \notag \\
&\rightarrow &\frac{1}{R}.  \label{14}
\end{eqnarray}%

Now from the Ellis wormhole optical metric for very large, but constant $R$, it follows that
\begin{eqnarray}\notag
\lim_{R\rightarrow \infty } \mathrm{d}t &=&\lim_{R\rightarrow \infty
}\left( R \sqrt{1+\frac{a^2}{R^2}}\right) \mathrm{d}\varphi\\
&\to & R \, \mathrm{d}\varphi.   \label{15}
\end{eqnarray}%

From the last two equations we find that  $
\kappa (C_{R})\mathrm{d}t= \mathrm{d}\,\varphi
$. Considering the GBT and choosing the line as  $r=b/\sin \varphi$, where $b $
is known as the impact parameter which gives the minimal
radial distance.  In that case we can easely find the deflecton angle to be computed as follows 
\begin{eqnarray}\label{16}
\hat{\alpha}=-\int\limits_{0}^{\pi}\int\limits_{\frac{b}{\sin \varphi}}^{\infty}K\mathrm{d}S.
\end{eqnarray}

If we substitute the result found in Eq. \eqref{10} then the
deflection angle $\hat{\alpha}$ can be found by solving the following integral
\begin{eqnarray}\label{17}
\hat{\alpha}&=&-\int\limits_{0}^{\pi}\int\limits_{\frac{b}{\sin \varphi}}^{\infty}\left(-\frac{a^2}{\left(r^2+a^2\right)^2}\right)\sqrt{\det \tilde{g}}\mathrm{d}r\mathrm{d}\varphi\\\nonumber
&= &\pi-2\,\text{EllipticK}(-\frac{a^2}{b^2})
\end{eqnarray}

Solving this integral, we find that the deflection angle is given in terms of the complete elliptic integral of the first kind.  Approximationg the solution in leading order terms we find
\begin{equation}
\hat{\alpha}\simeq \frac{\pi a^2}{4 b^2}-\frac{9 \pi a^4}{64 b^4}+\mathcal{O}(\frac{a^6}{b^6}).  \label{18}
\end{equation}

As we can see this result is consistent with Eq. \eqref{1}. However we note that only the leading term in the deflection angle given by Eq. \eqref{18} is correct, while the second term is not (see for example \cite{wh4}). This is due to the straight line approximation used in computing the integral \eqref{17},  even though  Eq. \eqref{17} gives an exact result for the deflection angle with the integration domain $\mathcal{D}_{\infty}$. In principle, therefore, the problem of second-order correction terms can be resolved, but this goes beyond the scope of this paper.

\section{Janis-Newman-Winnicour wormhole}
\subsection{Gaussian Optical Curvature}
Let us write the Janis--Newman--Winnicour (JNW) wormhole solution in the form found by Virbhadra as follows \cite{wh11}
\begin{eqnarray}\notag
\mathrm{d}s^2&=&-\left(1-\frac{2m}{r}\right)^{\gamma}\mathrm{d}t^2+\left(1-\frac{2m}{r}\right)^{-\gamma}\mathrm{d}r^2+r^2 \times \\
&&\left(1-\frac{2m}{r}\right)^{1-\gamma}\left(\mathrm{d}\theta^2+\sin^2\theta \mathrm{d}\varphi^2 \right),  
\end{eqnarray}
where $\gamma=M/m$, in which $M$ is the Arnowitt, Deser and Misner (ADM) mass. Note that $M$ is related to the
asymptotic scalar charge $q$ by $M^2=m^2-k q^2/2$, where $k>0$ and represents the matter-scalar field coupling constant \cite{wh7}. Setting $\gamma>1$ corresponds to the JNW wormhole. In a similar fashion as in the last section we consider the null case and the equilateral plane yielding the JNW optical metric
\begin{equation}\label{20}
\mathrm{d}t^{2}=\frac{\mathrm{d}r^{2}}{\left(1-\frac{2m}{r}\right)^{2\gamma}}+\frac{r^2 \mathrm{d}\varphi^2}{\left(1-\frac{2m}{r}\right)^{2\gamma -1}}.
\end{equation}%

The new variable  $r^{\star }$, and the new function $f(r^{\star })$, takes the following form
\begin{equation}
\mathrm{d}r^{\star }=\frac{\mathrm{d}r}{\left(1-\frac{2m}{r}\right)^{\gamma}},\,\,\,f(r^{\star })=\frac{r}{\sqrt{\left(1-\frac{2m}{r}\right)^{2\gamma -1}}}%
. 
\end{equation}%

The nonzero Christoffel symbols to the optical metric are found as
\begin{eqnarray}
\tilde{\Gamma} _{\varphi \varphi }^{r}&=&-\left( r- \left( 2\,\gamma+1 \right) m \right) ,\\
\tilde{\Gamma} _{r\varphi }^{\varphi}&=&{\frac {r- \left( 2\,\gamma+1 \right) m}{r \left( r-2\,m \right) }}.
\end{eqnarray} 

With the Gaussian optical curvature 
\begin{equation}
K=\frac{m \left( 2 \gamma m +m -2 r \gamma \right)}{r^2 \left(r-2m\right)^2}\left(1-\frac{2m}{r}\right)^{2\gamma},
\end{equation}
or, after we consider only the linear terms in $m$, we find
\begin{equation}\label{25}
K\approx - \frac{2m \gamma}{r^3} \left(1-\frac{2m}{r}\right)^{2\gamma}.
\end{equation}

\subsection{Deflection angle by JNW wormhole}

We can now finally  apply GBT to the JNW wormhole optical metric  by calculating first the geodesic curvature with the help of the nonzero Christoffel  symboles and the unit speed conition. In this particular case we find
\begin{eqnarray}
\lim_{R\rightarrow \infty }\kappa (C_{R})&=&\lim_{R\rightarrow \infty } \left\lbrace \frac{R-\left(1+2 \gamma \right)m}{R^2 \left(1-\frac{2m}{R}\right)^{1-\gamma}}\right\rbrace,  \notag \\
&\rightarrow &\frac{1}{R}.  \label{14}
\end{eqnarray}%
and similary  $\mathrm{d}t\to R \,\mathrm{d}\varphi$. Using the result \eqref{25} the deflection angle gives the following integral
\begin{equation}
\hat{\alpha}=-\int\limits_{0}^{\pi}\int\limits_{\frac{b}{\sin \varphi}}^{\infty}\left[- \frac{2m \gamma}{r^3} \left(1-\frac{2m}{r}\right)^{2\gamma}\right]\sqrt{\det \tilde{g}}\mathrm{d}r\mathrm{d}\varphi\\\nonumber.
\end{equation}

Evaluating the above integral we find the solution to be in terms of the Appell hypergeometric function  of two variables
\begin{widetext}
\begin{equation}
\hat{\alpha}=\frac{\gamma}{2 \gamma+1}\left[\pi-\frac{mb}{\sqrt{(2m+b)(2m-b)}}\frac{\text{AppellF1}\left(2(1+\gamma);\frac{1}{2},\frac{1}{2}; 3+2\gamma; \frac{b}{b+2m}, \frac{b}{b-2m}\right)} {1+\gamma} \right].
\end{equation}
\end{widetext}

In what follows we shall consider three special cases:
\begin{itemize}
\item First case: $\gamma=2$ 
\end{itemize}

Letting $\gamma=2$, corresponds to JNW wormhole  when the scalar charge $q$ is complex \cite{wh2}. In this particular case, the deflection angle is found to be 
\begin{equation}
\hat{\alpha}\simeq \frac{8 m}{b}-\frac{8 m^2 \pi}{b^2}.  
\end{equation}

\begin{itemize}
\item Second case: $\gamma=1/2$ 
\end{itemize}
Setting $\gamma=1/2$, corresponds to JNW wormhole when the scalar charge $q$ is real. The deflection angle is found to be
\begin{equation}
\hat{\alpha}\simeq \frac{2 m}{b}-\frac{ m^2 \pi}{2 b^2}.
\end{equation}

\begin{itemize}
\item Third case: $\gamma=1$ 
\end{itemize}
Setting $\gamma=1$, one finds the Schwarzschild spacetime. The result can be approximated as 
\begin{equation}
\hat{\alpha}\simeq \frac{4 m}{b}-\frac{2 m^2 \pi}{b^2}.  \label{30}
\end{equation}

Hence, by comparing these results with our Eq. \eqref{2} we conclude  that only the first order terms in $m$ agrees with the results reported in Ref. \cite{wh2}. Again, this is to be expected due to the straight line approximation, in other words the agreement between the GW method and the geodesics approach breaks down for the second order terms like $m^2$.

\section{Conclusion}

In this paper, we have calculated the deflection angle of light in a Ellis wormhole and Janis--Newman--Winnicour (JNW) wormhole geometry using the GW method by extending the GW method in the context of wormholes. We have introduced the JNW and Ellis wormhole optical metrics and employed the GBT to these metrics. Then we have found an \textit{exact} result in the weak limit approximation in leading order terms, unfortunately, this manuscript does not study the second-order corrections  due to the straight line approximation involved while computing the deflection angle. In the case of Ellis wormhole, we have shown that the deflection angle is given in terms of the complete elliptic integral of the first kind,  whereas in the case of JNW wormhole, the deflection angle is expressed in terms of Appell hypergeometric functions. The importance of these results on the the other hand, are purely conceptual, since the gravitational lensing phenomenon can also be viewed as partially topological effect of the global spacetime topology with a domain of integration \textit{outside} of the light ray.

\section*{ACKNOWLEDGEMENTS}
I would like to thank the anonymous referee and the editor for their valuable and constructive suggestions  to improve
the paper. Also I would like to thank Prof. Gerard Clement and Dr. Naoki Tsukamoto for their comments on my paper.

\end{document}